\documentclass{moriond}

\def\be{\begin{equation}}
\def\ee{\end{equation}}
\def\bea{\begin{eqnarray}}
\def\eea{\end{eqnarray}}

\newcommand{\Photo}{\includegraphics[height=45mm]{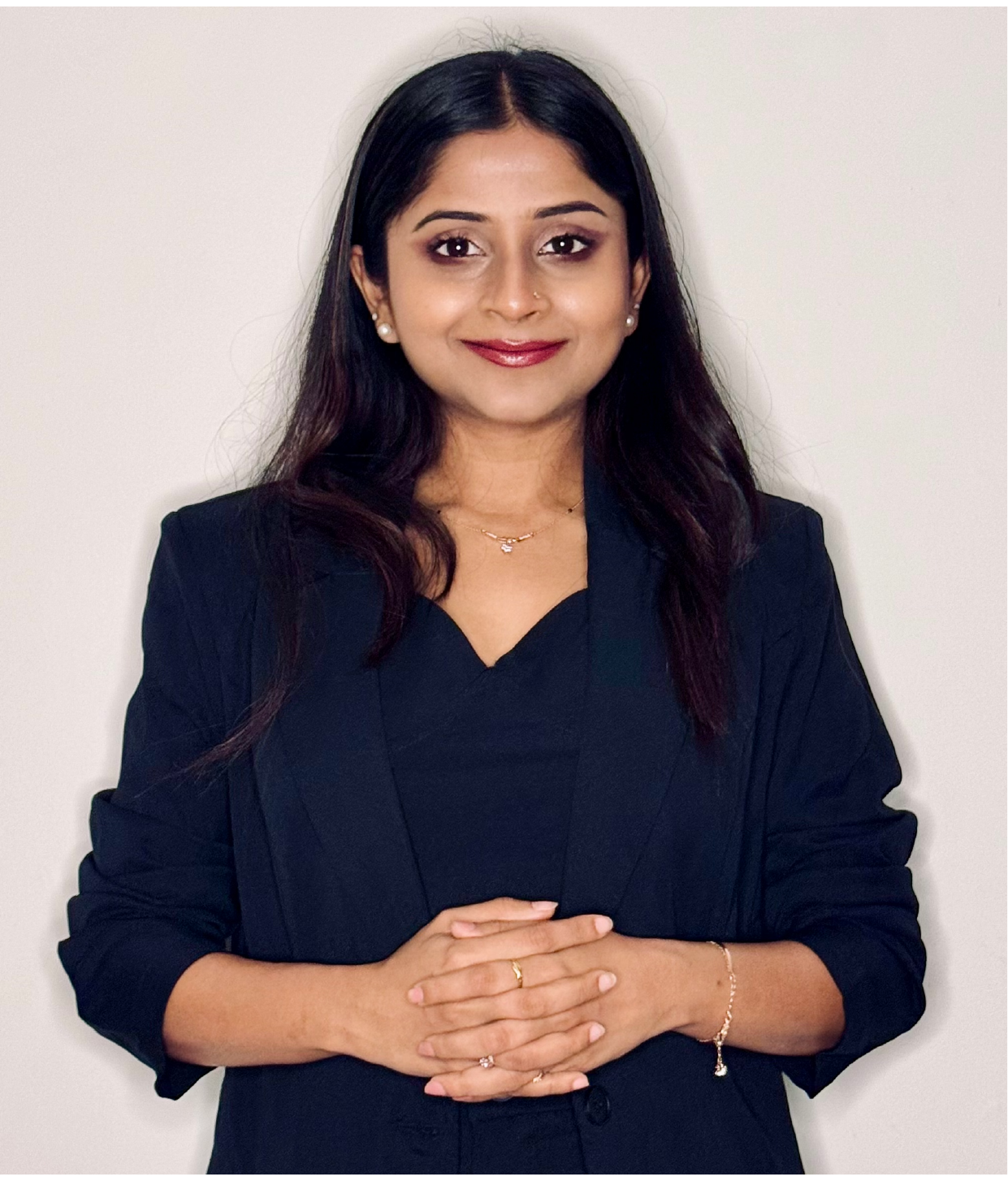}}
\usepackage{wrapfig}  
\usepackage{graphicx}
\usepackage{multirow}
\usepackage[square,numbers]{natbib}
\begin{document}
\vspace*{4cm}
\title{Charged-particle production in pp collisions at 13.6 TeV and Pb-Pb collisions at 5.36 TeV with ALICE}

\author{ Tulika Tripathy (for the ALICE collaboration) }

\address{Laboratoire de Physique de Clermont Auvergne, Université Clermont Auvergne, Clermont-Ferrand, 60026, France}

\maketitle\abstracts{
The first measurements of the charged-particle pseudorapidity density, ${\rm d}N_{\rm{ch}}/{\rm d}\eta$, in proton–proton (pp) collisions at $\sqrt{s} =$ 13.6 TeV and in lead–lead (Pb–Pb) collisions at $\sqrt{s_{\rm{NN}}} = $ 5.36 TeV are presented. The analysis is based on Run 3 data recorded in 2022 and 2023, using the upgraded ALICE detector at the LHC. The average charged-particle pseudorapidity density $\langle{\rm d}N_{\rm{ch}}/{\rm d}\eta\rangle$ in pp collisions was measured at midrapidity ($|\eta|<$0.5) with the upgraded central barrel detectors, such as the Inner Tracking System and the Time Projection Chamber 
and, at forward pseudorapidity ($-$3.6 $<|\eta|<-$ 2.4), using the newly installed Muon Forward Tracker (MFT). For Pb--Pb collisions, the measurement was performed in different centrality classes, ranging from the most central 0–5$\%$) to the peripheral (70–80$\%$) collisions.}
\vspace{-1em} 
\section{Introduction}
The study of charged-particle pseudorapidity density ($\mathrm{d}N_{\mathrm{ch}}/\mathrm{d}\eta$) at central and forward rapidities is crucial for exploring the dynamics of particle production in high-energy collisions. This observable provides insight into the initial conditions and energy density of the matter formed in such events.  At LHC energies, final-state particle production arises from the interplay between soft and hard QCD processes, influenced by non-linear QCD evolution in the initial state. Understanding how $\mathrm{d}N_{\mathrm{ch}}/\mathrm{d}\eta$ varies with collision system and energy helps disentangle these respective contributions~\cite{ALICE:2025cjn}.

Results from pp collisions at $\sqrt{s} = 13.6$ TeV and Pb--Pb collisions at $\sqrt{s_{\mathrm{NN}}} = 5.36$ TeV are presented using ALICE Run 3 data. The charged-particle density is measured at midrapidity ($|\eta| < 0.5$) and forward rapidity ($-$3.6 $<|\eta|<-$ 2.4), and is compared to prior data and theoretical models.

\section{Experimental Apparatus}
ALICE underwent a major upgrade during Long Shutdown 2 (2019–2022) of the CERN LHC. Key improvements include installation of the Fast Interaction Trigger (FIT), the Muon Forward Tracker (MFT), the upgraded Inner Tracking System (ITS2) and the Time Projection Chamber (TPC),  and an online–offline ($O^2$) data processing framework. The ITS2 consists of seven layers of ALPIDE monolithic active pixel sensors (MAPS), offering enhanced pointing and vertex resolution due to their proximity to the interaction point~\cite{ALICE:2023udb}. In the TPC, the previous multiwire proportional chambers were replaced with Gas Electron Multiplier (GEM) foils. The readout method was fully overhauled to enable continuous, non-triggered data acquisition from the core detectors. This is facilitated by FIT, which enables precise timing for continuous readout and is also used to estimate centrality, select events, and determine the collision time. The MFT is a silicon pixel tracker placed between the ITS and the muon absorber. It consists of two half-cones, each with five half-disks made of front and back detection layers, and offers a pointing resolution of a few tens of $\mu$m for forward muons. More details can be found in Ref.~\cite{ALICE:2023udb}.
\vspace{-1em} 
\section{Analysis details}
Minimum bias events are selected based on the relative time difference between the FIT time-zero (FT0) signals and the beam bunch crossing time, excluding events near the ITS readout frame edges or those with vertex mismatches between vertex estimations from FT0 and the combined TPC and ITS detectors exceeding 1 cm.  Only events with reconstructed vertices within $\pm$10 cm from the nominal interaction point are considered. Inelastic events with at least one charged particle within $|\eta|<1$ (INEL$>$0 event class) are used. Around 3 billion, 2.1 billion, and 1.1 billion events were accepted after the above event selection criteria in analyses of pp collisions at mid-, forward-pseudorapidity, and Pb--Pb collisions at mid-pseudorapidity, respectively. Centrality in Pb–Pb collisions is determined by fitting the FT0 amplitude ($-$3.3 $< \eta <$ $-$2.1) using a Glauber MC model~\cite{Loizides:2017ack}. Only global (matched between ITS and TPC) and ITS-only tracks are considered in pp and Pb-Pb collisions at mid-pseudorapidity, and tracks that pass through at least 5 layers of MFT are considered for pp collisions at forward rapidity (see Table~\ref{tab1}). Corrections for acceptance and tracking efficiency are applied using \textsc{Pythia}~8/Angantyr simulations. The charged-particle pseudorapidity distribution, or density, is obtained by counting the number of primary charged particles per collision within defined pseudorapidity intervals.
\begin{table} [!hpt]
\centering
                \begin{tabular}{|c |c | c|c |}

                \hline
                \multirow{2}{*}\textbf{ } & Pb--Pb collisions (Central)& \multicolumn{2}{c|}{pp collisions}  \\
\cline{3-4}
 &  & Central & Forward   \\
\hline

                $\eta$ coverage &  $|\eta| < 1.0$ & $|\eta| < 0.5$ &  $-$3.6 $<|\eta|<-$ 2.4   \\
                \hline 
                Track type & Global tracks & Global tracks & MFT tracks\\
                & + ITS only tracks &+ ITS only tracks & with hits on 5 layers
                \\

                \hline 
                \end{tabular}
               
               \caption{Summary of the track selections used in the analysis. Here, global tracks are the ones that are matched between ITS and TPC.}                
               \label{tab1}
\end{table}
\vspace{-1em}  
\section{Results}
\vspace{-0.9em} 
\begin{figure}[hpt!]
    \centering
    \includegraphics[width=0.6\textwidth]{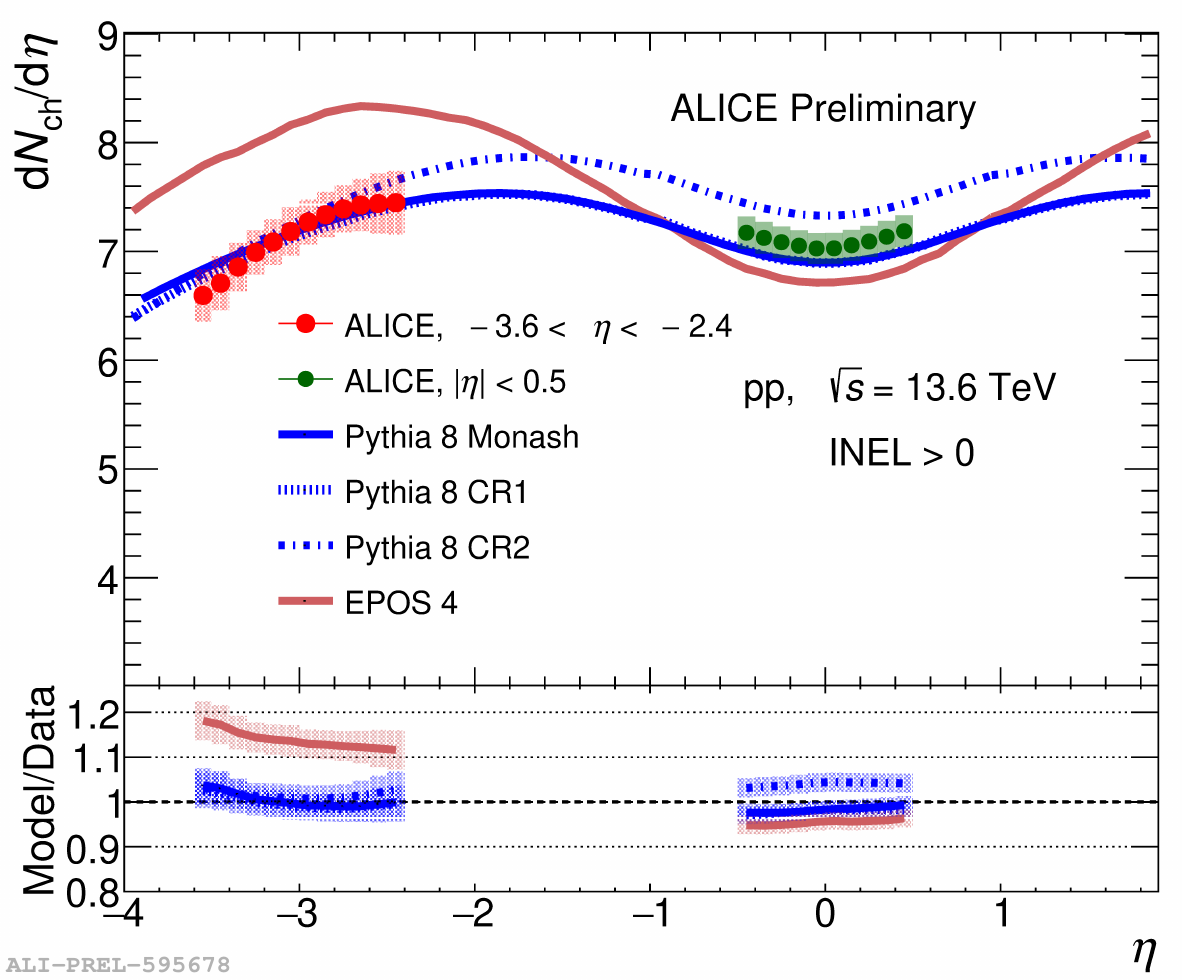} 
    \caption{Upper panel: Pseudorapidity dependence of the charged-particle pseudorapidity density for pp collisions at  $\sqrt{s}$ = 13.6 TeV at forward pseudorapidity (red solid circles) and at $\sqrt{s}$ = 13.6 TeV in mid-rapidity (green solid circles). The results are compared with EPOS4 and different tunes of \textsc{Pythia}\~8. Lower panel: the ratio of model to data.}
    \label{fig:1}
\end{figure}
\subsection{Results in pp collisions at $\sqrt{s} = 13.6 $ TeV}

Figure~\ref{fig:1} shows the pseudorapidity density distribution of charged particles in pp collisions for INEL$>$0 events for mid-pseudorapidity ($|\eta| < 0.5$) and in forward-pseudorapidity ($-$3.6 $<|\eta|<-$ 2.4) using MFT. The ALICE experiment is able to measure charged-particle multiplicity density with tracks in the central and forward rapidity region for the first time. It adds discrimination power when compared to models. The results are compared with calculations from \textsc{Pythia}~8 Monash tune~\cite{Skands:2014pea}, and two color reconnection (CR) modes. The \textsc{Pythia}~8 Monash tune includes an MPI-based CR scheme, while  CR1 mode includes the QCD-based CR scheme~\cite{Christiansen:2015yqa} and CR2 includes the gluon-move scheme~\cite{Bierlich:2022pfr}. The \textsc{Pythia}~8 Monash and CR1 tunes are consistent with the data within uncertainties in both rapidity regions, while CR2 overestimates the data at mid-$\eta$ by $\sim$ 4\%. The data are also compared with EPOS 4 model~\cite{Werner:2024ntd}, which includes a core that evolves hydrodynamically, and is surrounded by a more dilute corona for which fragmentation occurs. EPOS 4 overestimates the data at forward rapidity by $\sim$15\% and underestimates the data at mid-rapidity by $ \sim$5\%. 
\vspace{-1em}  
\subsection{Results in Pb--Pb collisions at $\sqrt{s_{\rm NN}} =$ 5.36 TeV}
\begin{figure}[hpt!]
    \begin{center}
    \includegraphics[width = 0.46\textwidth]
    {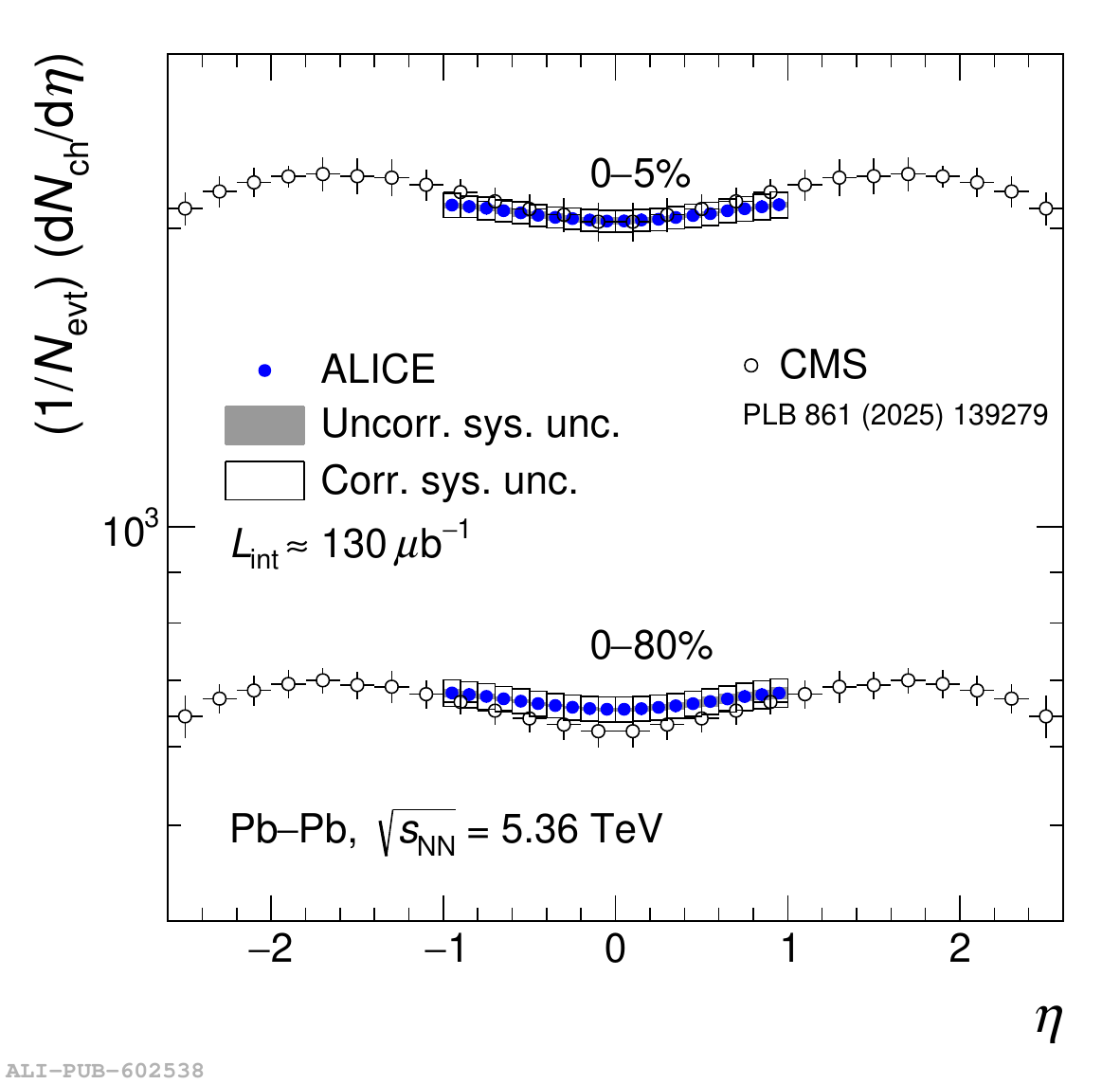}
      \includegraphics[width = 0.46\textwidth]
           {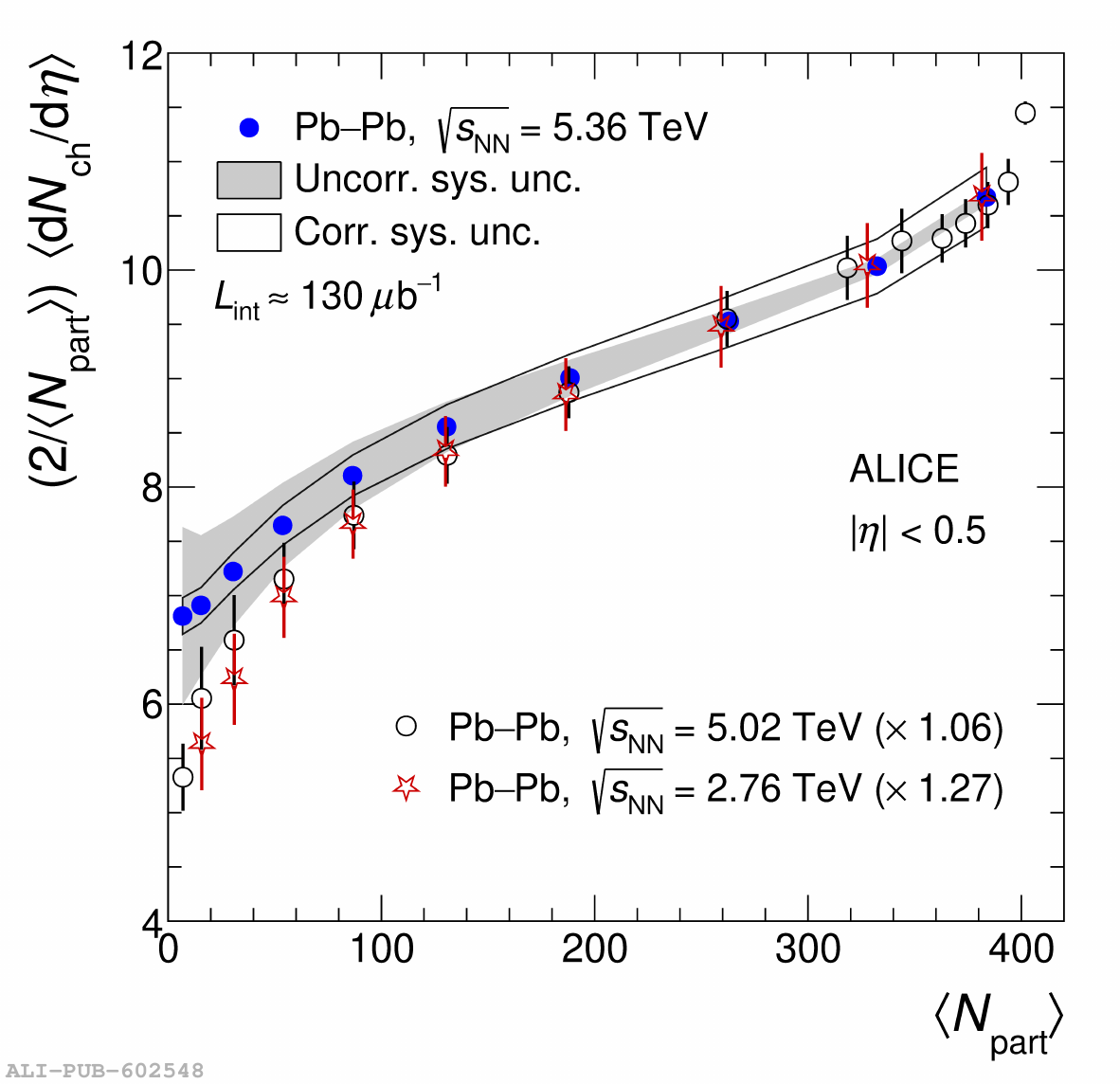}
    \end{center}
    \caption{Left: Comparison of charged-particle pseudorapidity density in 0–80\% and 0–5\% Pb–Pb collisions from ALICE ($|\eta| < 1$) and CMS ($|\eta| < 2.6$). ALICE data show uncorrelated (grey bands) and correlated (open boxes) systematic uncertainties; statistical uncertainties are negligible. CMS uncertainties are shown as vertical bars.
Right: Centrality dependence of $(2/\langle N_{\mathrm{part}} \rangle)\langle \mathrm{d}N_{\mathrm{ch}}/\mathrm{d}\eta \rangle$ in Pb–Pb collisions at $\sqrt{s\_{\mathrm{NN}}} = 5.36$ TeV. Grey bands and open boxes represent uncorrelated and correlated uncertainties. Scaled data from 2.76 TeV (×1.27) and 5.02 TeV (×1.06) are also shown for comparison.}
    \label{fig:3}
\end{figure}

Figure~\ref{fig:3} (left) presents a comparison of the charged-particle pseudorapidity density measured by ALICE in 0--80\% and 0--5\% central Pb--Pb collisions with recent results from the CMS experiment~\cite{CMS:2024ykx}. ALICE defines centrality based on charged-particle multiplicity, while CMS relies on transverse energy measured in forward calorimeters. The CMS measurements extend up to $|\eta| < 2.6$, compared to ALICE's midrapidity range. Despite these differences in methodology and pseudorapidity coverage, the results from both experiments are consistent within uncertainties.

Figure~\ref{fig:3} (right) shows the values of $(2/\langle N_{\mathrm{part}} \rangle)\langle \mathrm{d}N_{\mathrm{ch}}/\mathrm{d}\eta \rangle$ (solid blue circles) as a function of average number of participating nucleons, $\langle N_{\mathrm{part}} \rangle$, in Pb--Pb collisions at $\sqrt{s_{\mathrm{NN}}} = 5.36$~TeV. A clear centrality dependence is observed, with the normalized pseudorapidity density decreasing by a factor of approximately 1.8 from the most central (high $\langle N_{\mathrm{part}} \rangle$) to the most peripheral (low $\langle N_{\mathrm{part}} \rangle$) events. The new results are compared with previous Pb–Pb data at $\sqrt{s_{\mathrm{NN}}} = 2.76$~TeV~\cite{ALICE:2010mlf}, scaled by 1.27, and at $\sqrt{s_{\mathrm{NN}}} = 5.02$~TeV~\cite{ALICE:2015juo}, scaled by 1.06. These scaling factors correspond to the observed $\sqrt{s_{\mathrm{NN}}}^{0.156}$ dependence of multiplicity in central collisions~\cite{ALICE:2025cjn}. After scaling, the 2.76 and 5.02 TeV measurements are in good agreement with the 5.36 TeV data, within uncertainties and assuming full uncorrelation between data sets.

\section{Summary}
This proceeding presents measurements of charged-particle pseudorapidity density, $\langle \mathrm{d}N_{\mathrm{ch}}/\mathrm{d}\eta \rangle$, in pp collisions at $\sqrt{s}$ = 13.6 TeV and Pb-–Pb collisions at $\sqrt{s_{\mathrm{NN}}} = 5.36$ TeV using the upgraded ALICE detector during LHC Run 3. In pp collisions, measurements were performed at midrapidity ($|\eta|<0.5$) with ITS2 and TPC, and at forward rapidity ($-3.6 < \eta < -2.4$) using the newly installed Muon Forward Tracker (MFT). For Pb–Pb collisions, the analysis covers centrality intervals from 0–5\% (most central) to 70–80\% (peripheral).
In pp collisions, the measured $\langle \mathrm{d}N\_{\mathrm{ch}}/\mathrm{d}\eta \rangle$ values agree well with predictions from \textsc{Pythia}~8 and show consistent trends between mid- and forward-rapidity regions. Model comparisons reveal differences in predictions (EPOS 4 vs Pythia 8 vs data) at forward $\eta$.
In Pb–Pb collisions, a strong centrality dependence of charged-particle production is observed. Charged-particle multiplicity normalized by $\langle N_{\mathrm{part}} \rangle$ decreases by a factor of nearly 1.8 from central to peripheral events. Comparisons with scaled data at $\sqrt{s_{\mathrm{NN}}} = 2.76$ and 5.02 TeV confirm consistency with the new 5.36 TeV measurements, supporting the established energy scaling trend. The results are also in agreement with those from the CMS experiment despite differences in centrality and pseudorapidity definitions.


\end{document}